\documentclass[12pt,preprint]{aastex}

\usepackage{amsmath}
\usepackage{txfonts}
\usepackage{graphicx}
\usepackage{epstopdf}
\usepackage{natbib}
\usepackage{color}
\newcommand{\PM}{\emph{PM}}

\newcommand{\otma}{OTMA}

\newcommand{\utma}{UTMA}

\newcommand{\done}{$d_1$}

\newcommand{\Dthree}{$D_3$}

\newcommand{\fpri}{$F_{pri}$}
\newcommand{\ftwo}{$F_2$}
\newcommand{\fthree}{$F_3$}
\newcommand{\Fp}{$F_p$}

\newcommand{\Yone}{$Y_1$}

\newcommand{\uzero}{$\bar{u}_0$}

\shorttitle{Obstructed vs Unobstructed Telescopes}
\shortauthors{Balasubramanian S}

\begin{document}


\title{Obstructed Telescopes vs Unobstructed Telescopes for Wide Field Survey -- A Quantitative Analysis}

\author{Balasubramanian Singaravelu\altaffilmark{1,2} and Remi A. Cabanac\altaffilmark{1,2}}
\affil{ Universit\'{e} de Toulouse 3 Paul Sabatier;  Toulouse, France\\}

\altaffiltext{1}{Universit\'{e} de Toulouse 3 Paul Sabatier;  Toulouse, France}
\altaffiltext{2}{CNRS; IRAP; 57, Avenue d'Azereix, BP 826, F-65008 Tarbes cedex, France}

\begin{abstract}
Telescopes with unobstructed pupil are known to deliver clean point spread function (PSF) to their focal plane, in contrast to traditional telescopes with obstructed pupil. Recent progress in the manufacturing aspheric surfaces and mounting accuracy favors unobstructed telescopes over obstructed telescopes for science cases that demand stable and clean PSF over the entire field--of--view. In this paper we compare the image quality of an unobstructed Three--Mirror--Anastigmat (TMA) design with that of an obstructed TMA. Both the designs have the same primary mirror, effective focal length, field--of--view and detector characteristics. We demonstrate using simulated images of faint elliptical galaxies imaged through the two designs, that both the designs can measure morphological parameters with same precision,  if the PSF is reconstructed within 12 arc-minutes of the source. We also demonstrate that, the unobstructed design delivers desirable precision even if the PSF is reconstructed 50 arc-minutes away from the source. Therefore the PSF of unobstructed design is uniform over a wider field--of--view compared to an obstructed design. The image quality is given by the 1$\sigma$ error-bars (68\% confidence level) in the fitted values of the axis--ratio and position--angle of the simulated galaxies.
\end{abstract}


\section{Introduction} 
Many astrophysical observations demand very high-resolution imaging and a stable PSF over the desired field--of--view. The former is limited by the size of the pupil and presence of central obstruction, while the latter is limited only by the obstruction. Theoretically unobstructed pupils deliver a simpler PSF compared to obstructed designs. We know from the studies of \citet*{row10} that the accurate knowledge of the PSF is significant to make precise science measurements on astronomical images.  Recent works done by \citet*{lam10} and \citet*{lev11} suggest that unobstructed pupil have a faster survey rate for wide-field survey missions to study dark energy. These wide-survey images are used to perform strong lensing and weak lensing analyses. We know that lensing analyses depends on the precision with which we can measure the ellipticity and position--angle of a large number background sources and hence the shear and convergence produced by the lensing mass. The objective of this paper is to quantify the image quality of an obstructed and unobstructed telescope for such a wide-field survey mission. In this paper we focus on space based telescopes because ground based telescopes even with `multi conjugate adaptive optics' (MCAO) are still not suitable to perform high resolution surveys. 

\subsection{Three Mirror Anastigmat Telescopes}

Optical designs based on three aspheric mirrors (dubbed Three--Mirror--Anastigmat or TMA) are favored in modern astronomical instrumentation because these telescopes have a wider field--of--view for a given pupil size. And it is possible to build a range of diffraction-limited designs using this configuration. TMA telescopes were proposed by \citet*{pau35} and developed by \citet*{bak69} and \citet*{kor72}. Traditionally, these telescopes consist of a secondary mirror which partially obstructs the light falling on the primary mirror. This type of three mirror telescope are called obstructed TMA (hereafter OTMA). A few telescopes have been built recently in which the secondary mirror is offset from the path of the incoming light and no obstruction is present in front of the primary reflector. These type of telescopes are called unobstructed TMA (hereafter UTMA). The design of UTMA telescopes was discussed by \citet*{coo79} and \citet*{kor80}. For stellar astrometry, UTMA design delivers better precision, for example GAIA \citep*{per05}. Also unobstructed telescopes can image exoplanets close to bright stars with high contrast \citep*{ser10}. Hence, UTMA designs have been proposed for exoplanet characterization missions like EChO \citep*{tin12} and SPICES \citep*{boc12}. In the next section we briefly summarize the theoretical advantage of using an unobstructed pupil.

\subsection{Advantage of an unobstructed pupil}

The effective light gathering area of an obstructed pupil is lesser than an unobstructed pupil. And the blur size on the image plane increases with increasing obstruction. The focal image of a point source at infinity or PSF of a focusing optical system is the Fourier transform of its pupil shape. In the case of a circular pupil, the shape of the PSF is the Airy pattern with a bright central disc and faint concentric rings. We know from the studies of \citet*{tay58} that in the Airy pattern of an annular aperture there is significant transfer of energy from the central disc to the outer rings. The presence of support structure will also produce additional artefacts on the PSF. 

\citet*{lev11}  show in their work that the survey rate for a mission is directly proportional to the light gathering area and inversely proportional to the blur size for a given SNR. Therefore unobstructed telescopes must have faster survey speed for a given SNR. They also demonstrate that for unobstructed telescopes there is an increased density of resolved galaxies for weak lensing analyses.  \citet*{lam10} show that the diffraction pattern of an OTMA telescope can destroy or mimic the lensing signals we desire to study because of the larger blur size. Hence the tighter PSF of UTMA telescopes should be beneficial for wide-field lensing missions. In this work we will quantify the precision with which UTMA telescopes can perform science measurements for weak lensing analyses compared to OTMA telescopes having same characteristics.

\subsection{Outline}
The objective of this work is to quantify the precision with which UTMA and OTMA telescopes can perform  morphological parameter measurements for a wide field survey. In order to reach this goal the work was organized as follows. First, we modeled and optimized in parallel an OTMA and an UTMA telescope, both having the same primary mirror, effective focal length and FoV. The science requirements for our study are same as that of the {\it Euclid} mission \citep*{euc11}. Second, we created an end-to-end semi-realistic samples of elliptical galaxies passed through the full instrumental path, PSF convolution, CCD pixelisation, and noise effects. Third we selected and controlled the biases of the fitting routine for measuring galaxy morphologies. Then we measured the ellipticity in the simulated galaxies and calculated the error introduced by the PSF on the error budget. Finally, we compare the precision with which both the designs can perform the desired measurements. The paper structure sequentially follows these steps.

%__________________________________________________________________

\section{Telescope design and optimization}
To compare the image quality of the two designs we need astronomical images observed using both the designs. To simulate quasi-realistic images we need atleast the PSF of the designs over the desired field--of--view (FoV). The end goal of this section is to compute the PSF for optimized designs of both the telescopes.

In this section, we present an overview of the method we used to design and optimize the OTMA and UTMA telescopes. First, we recall the parameters and conventions which are used to describe a telescope. Next, the procedure to derive first-order telescope design parameters is detailed.  Finally, using the telescope design parameters we design and optimize the telescopes in a commercially available optical design program \textsc{zemax}\footnote{www.zemax.com}. We use the same program to compute the spot diagram, EE plots and PSF at the desired locations over the desired field--of--view (FoV).

From first principles, we recall that due to aberrations and diffraction there is never a one--to--one correspondence between a source and its image on the focal plane. Geometrical optics, which treats light as a collection of rays traveling in straight lines is a good starting point to reduce the blur size of the focal image in the desired FoV . A design is successful when the optical aberrations it produces are small compared to the absolute physical limits set by diffraction. This abberation limited system will be further optimized using \textsc{zemax}, which takes into account the diffraction effects.

\subsection{Telescope Design Parameters}

\begin{figure}
\begin{center}
\includegraphics[width = 3in, keepaspectratio=true]{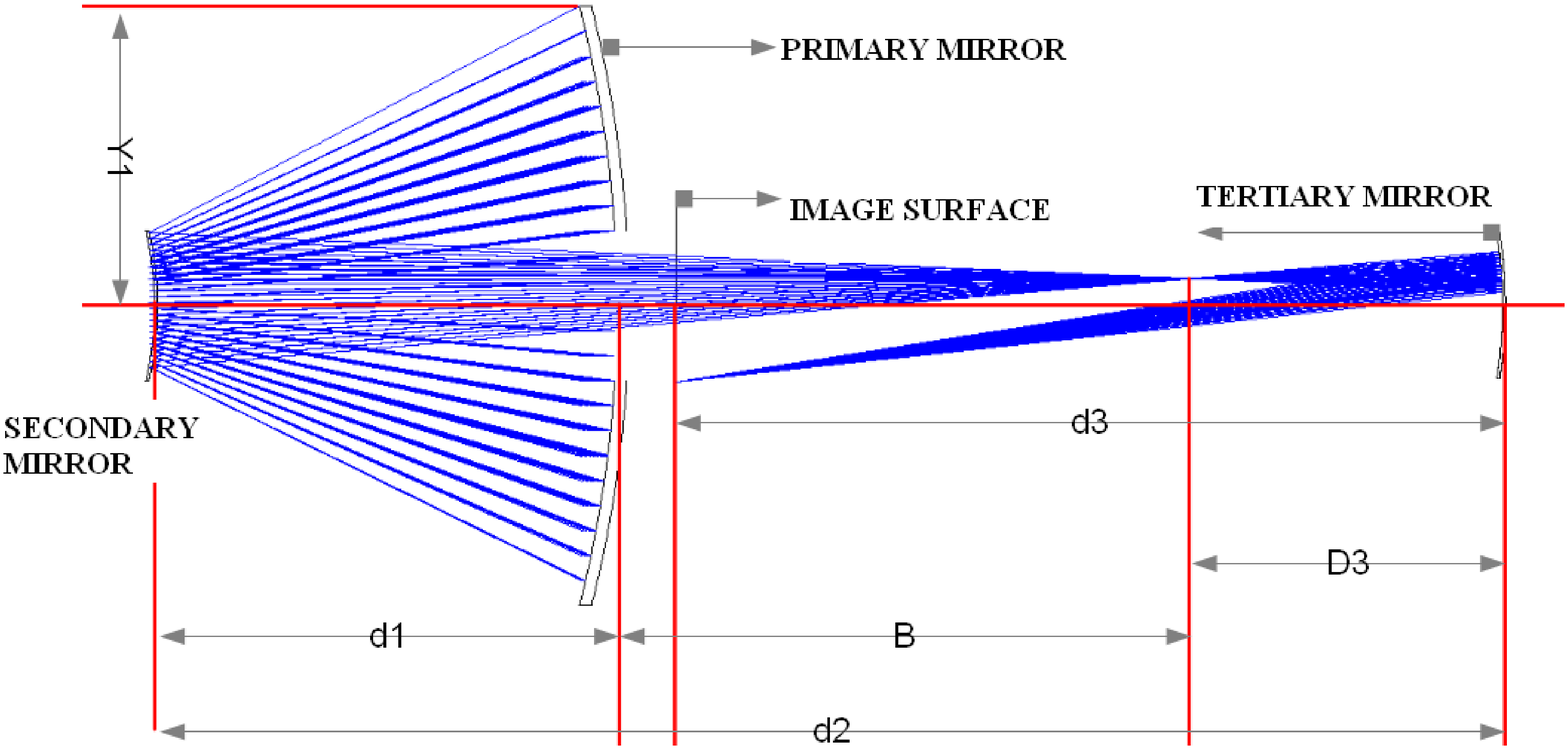}
\caption {{\bf Nomenclature for a TMA}. Figure showing the naming convention of the fundamental and constructional parameters associated with TMA telescopes in this work (Robb 1978 convention is followed)}
\label{tmaparam}
\end{center}
\end{figure}

By definition, Three--Mirror--Anastigmat (TMA) telescopes consists of three conic reflecting surfaces (see Figure \ref{tmaparam}). The characteristics of a TMA can be defined using two families of parameters. The fundamental design parameters, which define the optical configuration from an image perspective and the constructional parameters, which define the optical configuration from an engineering perspective. The fundamental design parameters constrain the pixel scale, resolution limit, FoV, etc. The fundamental design parameters as defined in \citep*{rob78}  are given below.

\begin{description}
\item [$F_3$] Focal length of the three mirror system, always positive
\item [$Y_1$] Height or radius of the primary mirror (PM)
\item [$F_{pri}$] Focal ratio of the  PM 
\item [$F_2$] Focal length of the two mirror system, set to be positive for Cassegrain type, and negative for Gregorian type
\item [B] Location of the two mirror focus with respect to the vertex of the PM, positive if beyond the vertex of the PM, negative if inside.
\item [$D_3$] Location of the tertiary mirror (TM) with respect to the two mirror focus, positive if beyond the focus, negative if inside.
\item [$\bar{u}_0$] Slope of the paraxial chief ray entering the system
\end{description}

The constructional parameters allow us to construct the telescope or to model them in an optical design program. Constructional parameters of TMAs constitute the surface curvatures $\{c_1, c_2, c_3\}$, conic constants $\{k_1, k_2, k_3\}$ and the separation between the surfaces $\{d_1, d_2, d_3\}$. A few of these parameters are shown in Figure \ref{tmaparam}. In addition to these constructional parameters UTMA telescopes have three decenter distances called $\{h_1,h_2,h_3\}$ that correspond to offset of the centre of the reflecting surfaces from the axis perpendicular to the focal plane. 

\subsection{Design Constraints}
The mission requirements fix the fundamental design parameters of the telescope.  We use the {\it Euclid} mission requirements \citep*{euc11} to fix the following parameters \{$Y_1=0.6$ m, $F_3 = 24.5$ m, FoV$= 0.787 \times 0.709$ deg$^2$\}. For {\it Euclid} the proposed wavelength range for the visual band instruments is 550--900 nm. Since it is not possible to introduce a spectrum in the optical design tool, we chose 3 discrete wavelengths to represent this band, they are \{550 nm, 725 nm, 900 nm\}. The diffraction effects were calculated for these wavelengths.

The fixed values of \{$Y_1$,$F_3$,FoV\} will be used as starting point for both the OTMA and UTMA designs. For wavelength 550 nm and $F_3 = 24.5$ m we obtan an image scale of 8.47 arc-sec/mm. For $Y_1=0.6$ m, we compute an Airy disk radius of 13.69 $\mu m$ (or 0.12\arcsec) for $\lambda\,=$ 550 nm and 22.41 $\mu m$ (or 0.19\arcsec) for $\lambda\,=$ 900 nm. The image scale and Airy radius dictate theoretical ability to detect PSF anisotropies close to the resolution limit. If the pixel size is larger than the Airy disk size, the PSF effects will not be noticeable. Therefore for our image quality analysis in Section 3 we simulate images with pixel scale 0.025\arcsec and 0.1\arcsec. The former pixel scale is much smaller than the Airy radius and the latter is in the order of the Airy radius. For pixel scales larger than this the OTMA and UTMA will have similar image quality. Also sources with artifacts smaller than the resolution limit are unresolved for both the designs.

\subsection{Obstructed TMA design}
The constructional parameters required to model an OTMA telescope can be obtained from the fundamental parameters using the relations derived in \citet*{rob78} (cf APPENDIX A for details). To calculate the constructional parameters we need all seven fundamental parameters. Three of the fundamental design parameters are constrained by the design requirement, so we have freedom to select the remaining parameters according to our practical constraints. In this work, we constrained the physical size of the designs and imposed a flat focal plane to obtain the remaining fundamental parameters (cf APPENDIX A for details). We generated a large family of OTMA telescope designs all obeying our constraints, from which one design was arbitrarily chosen to serve as a typical OTMA telescope. This telescope is named OTMA1. The fundamental design parameters and the constructional parameters for the chosen OTMA telescope are shown in Table \ref{tmaFundOtma} and Table \ref{tmaConstOParam} respectively. The central obstruction for the chosen design is 248 mm. Therefore, the linear obscuration for the design is 20.6\%.

\subsection{Unobstructed TMA design}

The design of UTMA telescopes were discussed by \citet*{coo79} and \citet*{kor80}. The work by \citet*{coo79} illustrates two examples of UTMA, but no analytical or numerical methods are provided to obtain those designs. The work by \citet*{kor80} gives a rigorous numerical technique to determine the shape of the mirror surfaces. Though, this method is very general it has the caveat of not being straightforward to implement with modern ray-tracing software (because this method uses non-conic surfaces) and is optimized only for the central FoV. In contrast, we propose here a simple method to design an UTMA telescope from an OTMA telescope.

\subsubsection{The Procedure To Design An UTMA Telescope From An OTMA Telescope}

\begin{figure*}
\begin{minipage}{7in}
\begin{center}
\includegraphics[width = 7in, keepaspectratio=true]{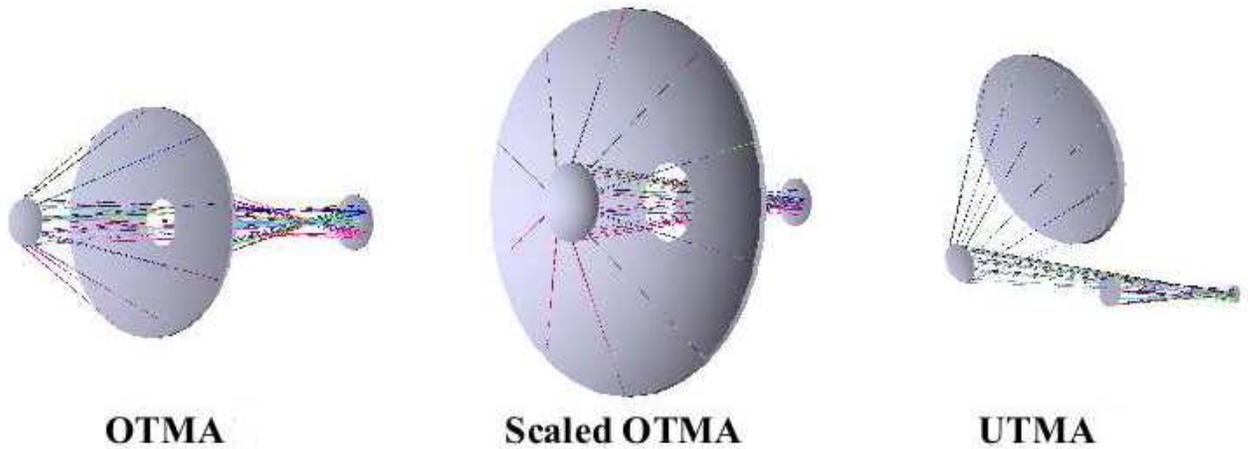}
\caption{{\bf UTMA design procedure.} A step-by-step pictorial representation for designing an UTMA telescope from an OTMA telescope is shown in this figure. Left panel is the initial OTMA1 telescope. Central panel is a scaled-OTMA telescope (here, scale factor $s \approx 2.5$). Right panel is a sub-part of the scaled telescope that can be used as a stand-alone UTMA telescope. The fundamental parameters for all the telescopes are given in Table \ref{tmaFundOtma}. The lines correspond to the light rays traced.}  
\label{uotmaSTEP}
\end{center}
\end{minipage}
\end{figure*}

The basic idea of the procedure is illustrated in Figure \ref{uotmaSTEP}. We select an unobstructed sub-pupil from the primary mirror (PM) of a given OTMA telescope. This unobstructed portion can be used as a stand-alone UTMA telescope. But the resulting telescope will have a smaller PM compared to the original OTMA telescope. To obtain an UTMA telescope with the desired aperture, we scale the OTMA telescope before separating a sub-pupil. We use the paraxial ray-trace equations and the condition that no rays incident on the primary surface should be obstructed by the secondary surface and tertiary surface (no vignetting) to obtain the `scale-factor'. This ensures us that the UTMA pupil is uniformly illuminated. 

To obtain the scale factor, let $y_j,d_j,i_j,u_j$ be the paraxial ray height, separation from the next surface, angle of incidence and angle after reflection. We already defined $Y_1$ to be the height of the PM, let the height of the scaled mirror be $Y_1'(=sY_1)$,  where $s$ is the scale-factor. Similarly, we denote the height of the secondary mirror (SM) and the scaled SM by $Y_2$ and $Y'_2(=sY_2)$. Using the ray trace equations in \citet*{bor65} we obtain the scale-factor as follows.

\begin{eqnarray}
y_j &=& y_{j-1} + d_{j-1}u_{j-1} \\
i_j &=& y_jc_j + u_{j-1} \\
u_j &=& u_{j-1} - 2i_j \\
y_2 &=& y_1 + d_1u_1 \\
u_1 &=& -2i_1  \\ \nonumber
&=& -2y_1c_1 \\
y_2 - y_1 &=& d_1(-2y_1c_1) \\
Y_1'-Y_2' &=& 2Y_1'c_1d_1 \\
\text{aperture of the UTMA} &=& 2sY_1c_1d_1 \\
s &=& \frac{\text{aperture of the UTMA}}{2Y_1c_1d_1}
\end{eqnarray}

We find that scaling does not change constructional parameters, but it does change  fundamental design parameters. Keeping the PM focal length ($F_p = 2F_{pri}Y_1 = 2F_{pri}'Y_1'$) invariant requires us to adjust the PM focal ratio. In other words, scaling $Y_1$ by a factor {\it s} decreases $F_{pri}$ by the same factor. Following our mission constraints, we applied this procedure on the OTMA1 design to obtain the scaled--OTMA1 design whose fundamental design parameters are shown in Table \ref{tmaFundOtma}. The scale-factor calculated using Equation (9), is approximately 2.5. The constructional parameters of the scaled design are unchanged. Unfortunately, this family of designs have an intrinsically narrow diffraction-limited  FoV. Here we derive a FoV of 0.3$\times$0.2 deg$^2$ (computed using \textsc{zemax}). The main reason for this narrow FoV is the decrease in $F_{pri}$ associated with the scaled design. The UTMA1 design is not suitable for our science requirement because of its narrow FoV, but science requirements less demanding in field size may find in this design an attractive and compact solution. 

A convenient way of increasing the UTMA FoV, is to start with an OTMA having a large $F_{pri}$. This is done at the expense of compactness of the final design. Hence, relaxing the constraint on size allows us to create a new family of OTMA designs (cf APPENDIX A). Again, one design (named OTMA2) is drawn randomly from this new family. Its fundamental parameters are given in Table  \ref{tmaFundOtma}. This design is scaled by scale-factor 2.6 to obtain the UTMA2 design. The constructional parameters for UTMA2 are shown in Table \ref{tmaConstOParam}. The OTMA1 and UTMA2 are chosen for optimization and image quality analysis. 

\begin{table}
\begin{center}
\caption{Fundamental design parameters of the chosen designs}
\begin{tabular}{lllllll}
\hline\hline
\noalign{\smallskip}
Design &$Y_1$[m] & $F_{pri}$  & $F_2$[m] & B[m] & $D_3$[m] \\
\noalign{\smallskip}
\hline
\noalign{\smallskip}
OTMA1 &0.60 & 1.00  &10.25 & 0.96 & 0.46 \\
scaled \otma1 &1.50  &0.40 &10.25 & 0.96 & 0.46 \\
\otma2 &0.60 &4.50 & 22.00 & 0.99 & 4.58 \\
scaled \otma2 &1.56 &1.73 & 22.00 & 0.99 & 4.58 \\
\noalign{\smallskip}
\hline
\end{tabular}
\label{tmaFundOtma}
\end{center}
$F_3 \, =$\, 24.5 m for all the above\\
scaled \otma\, is used as \utma
\end{table}

\begin{table}
\begin{center}
\caption{Constructional parameters of OTMA and UTMA telescopes}
\begin{tabular}{lccc}
\hline\hline
\noalign{\smallskip}
Parameter$^\dagger$ & OTMA1 & UTMA1 & UTMA2\\
\noalign{\smallskip}
\hline
\noalign{\smallskip}
$c_1$  & $-$1/2399 & $-$1/2399 & $-$1/10800\\
$c_2$ &  $-$1/513 &  $-$1/513 & $-$1/3339\\
$c_3$ & $+$1/652 & $+$1/652 &  $-$1/4834\\
$d_1$$^*$  & $-$973  & $-$973 & $-$4140\\
$d_2$$^*$ & $+$2038 & $+$2038 & $+$9722\\
$d_3$$^*$ & $-$1105 & $-$1105 & $-$5108\\
$k_1$ & $-$0.995 & $-$0.995 & $-$0.929\\
$k_2$ & $-$1.561 & $-$1.561 & $-$2.108\\
$k_3$ &  $-$0.756 &  $-$0.756 & $-$0.425\\
$h_1$$^*$ & na & 900  & 1050\\
$h_2$$^*$ & na & 180  & 267\\
$h_3$$^*$ & na &  0  & 0\\
SM diameter$^*$ & 246 &  246  & 331\\
TM diameter$^*$ & 276 &  150  & 876\\
\noalign{\smallskip}
\hline
$^\dagger$ {\footnotesize cf section 2.1 for explanation}\\
$^*$ {\footnotesize in units of mm}
\end{tabular}
\label{tmaConstOParam}
\end{center}
\end{table}

\subsection{Optimization}
\begin{figure}
\begin{center}
\includegraphics[width = 3in, keepaspectratio=true]{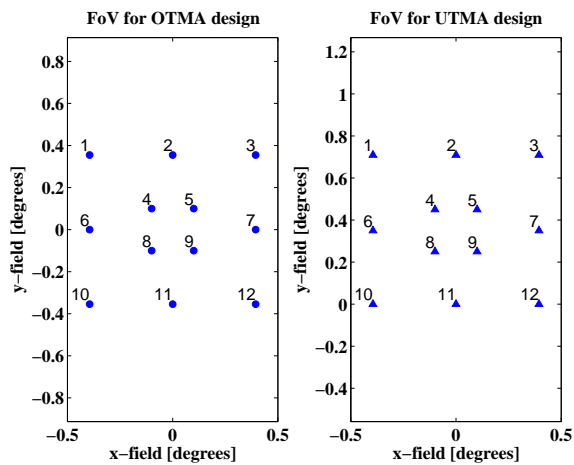}
\caption{{\bf Field of View.} The FoV points for the chosen OTMA and UTMA telescopes are shown. Both the designs have a FoV of $ 0.787 \times 0.709$ deg$^2$. For the obstructed design rays with angle \uzero\, and -\uzero\, are brought to focus. To achieve the same field of view in an UTMA design rays with \uzero\, and 2\uzero\, are brought to focus.}  
\label{tmaFoV}
\end{center}
\end{figure}

Using the constructional parameters in Table \ref{tmaConstOParam}, the two telescope designs are modeled in \textsc{zemax}. The optimization feature of \textsc{zemax} has the capability of improving the design given a reasonable starting point and a set of variable parameters. In our case, the starting point is the diffraction-limited system defined by the constructional parameters in Table \ref{tmaConstOParam} and the set of parameters to be optimized are all the constructional parameters plus the aspheric coefficients that define the mirror surfaces. In order to assess quantitatively how closely the optimized optical system meets our specified set of constraints, a merit-function is defined. The merit-function value is the square root of the weighted sum of the squares of the difference between the actual and desired value of the list of constraints. Here, we fixed the effective focal length ($F_3$) of the system and constrained the size of spot radii to give diffraction-limited images over the desired FoV. During optimization process \textsc{zemax} takes into account the wave nature of light and hence the diffraction effects. The spot diagrams were computed at locations given in Figure \ref{tmaFoV} to verify whether the constraints were met. Optimized OTMA and UTMA designs achieve similar Merit Function values. Next we compute the PSFs for both the optimized designs.

\subsection{PSF Computation}
\begin{figure}
\begin{center}
\includegraphics[width = 3in, keepaspectratio=true]{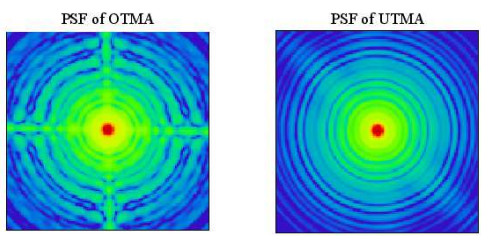}
\caption{{\bf PSF} A sample PSF for both the OTMA and UTMA designs at FoV location labeled `1' in Figure \ref{tmaFoV}. The OTMA PSF on the left panel shows diffraction spikes due to the presence of spider support structure. The UTMA PSF on the right panel is rotational invariant and any artifact visible to naked eye is due to the graphics file format.}  
\label{tmaPSF}
\end{center}
\end{figure}

The PSFs of OTMA and UTMA designs were computed using \textsc{zemax} (which employs Huygens-Fresnel diffraction principle) at field locations labeled in Figure \ref{tmaFoV}. The PSF computation was done for wavelength bands between 550 nm and 900 nm. The PSFs computed at these locations are converted into FITS images and normalized to unit intensity. Figure \ref{tmaPSF} shows the PSF for OTMA and UTMA telescope at FoV location labeled `1' in Figure \ref{tmaFoV}. The PSFs have a pixel scale of 0.025\arcsec (same as the value used for simulating galaxy images). At first glance, we can see that the the UTMA PSF on the right panel (Figure \ref{tmaPSF}) is rotational invariant and does not show any features like spikes compared to the OTMA PSF. These PSFs will be used to simulate galaxy images for the two designs.

Using the simulated images, our objective is to find out whether the presence of spikes and the brightness of the outer rings affect the science measurements of the OTMA telescope. The variation of PSF over the FoV and the PSF reconstruction errors will also affect the science measurements.  If the PSF is homogeneously uniform over the entire FoV, PSF reconstruction errors will be smaller. The intrinsic characteristics of the PSF and the variation of the PSF over the FoV will be addressed in section 4.

%__________________________________________________________________

\section{Image Simulation And Model Fitting}

In this section we brief on how to simulate quasi realistic images which can be obtained from the two designs. Then we show how to use model fitting tools on these images to fit the simulated morphological properties. In this study the PSF is the only distinguishing factor between the two designs. We add noise to the images to account for other dominating factors during model fitting. The remaining factors are assumed to affect the precision of both the designs in the same way. 

The procedure for image simulation and model fitting is as follows. First we simulated galaxy images with no PSF artifacts and free of any noise. Second we convolved the images with both OTMA and UTMA PSF to include the PSF effects. The simulated galaxy images and the PSF have same pixel scale of 0.025\arcsec. After convolution we pixelised the images to 0.1\arcsec pixel scale. This pixel scale is in the order of the Airy radius and allow us to study the effect of the PSF on different scales. We now have two sets of images with different pixel scales (or sampling frequencies). We add poissonian (sky background) noise to the images with both pixel scales. Finally, for measuring the axis--ratio and the position--angle we used the routine \textsc {sextractor} \citep*{ber96} to get the initial values, and then fit the same using 2D--parametric model fitting technique \textsc{galfit} \citep*{pen10}. 
% The sky background noise is a poissonian noise which is a function of sky background value (in ADU) and exposure time in seconds. This noise is considered to be the dominant noise for wide field survey.

\subsection{Simulating Galaxy Images}
Our galaxy images were simulated using the ASTROMATIC\footnote{www.astromatic.net} tools provided by \citet*{ber09}. These tools allow us to produce galaxy images with known properties (namely apparent magnitude, redshift, axis--ratio, position--angle, bulge-to-total ratio in the observed passband) and a de--Vacouleur profile. First, we used the tool \textsc{stuff} \citep*{ber09} to create a catalog of sources. These sources are faint elliptical galaxies with apparent magnitude between 23 mag and 24 mag (the {\it Euclid} mission is magnitude limited at 24.5 mag in visual band). \citet*{jou09} show that the number density of resolved objects is directly proportional to the resolving power of the telescope for a given magnitude limit.  The Airy radius of both our designs are in the same order (0.12\arcsec at 550nm and 0.19\arcsec at 900nm) and this limits the half-light radius for our simulated galaxies. We simulate galaxies with mean half-light radius in the order of 0.2\arcsec. 

The inputs for \textsc{stuff} routine are detailed Table \ref{stuff}. The catalog contains a large number of galaxies with a wide range of axis--ratios. For simplicity, we divide the sample into 8 bins according to the galaxy axis--ratio. Table \ref{galaxyBins} shows the bins, they are labeled E1 through E8. PSF convolution is scale dependent and affects different axis--ratios in different ways. In real surveys, galaxies with slender axis--ratios are rarer than galaxies which tend to be circular.  We ensure that there are at least hundred members in each galaxy bin.

\begin{table}
\begin{center}
\caption{Essential input information for \textsc{stuff} catalog generator}
\resizebox{8cm}{!}{
\begin{tabular}{ll}
\hline\hline
\noalign{\smallskip}
Input parameter & Value \\
\noalign{\smallskip}
\hline
\noalign{\smallskip}
Range of apparent magnitudes & 23-24 \\
Pixel size (arcsec) & 0.025 \\
Effective collecting area ($m^2$) & 78.53$^{[1]}$ (corresponds to 10m aperture) \\
Cosmological parameters & \{$\Omega_M = 0.3, \Omega_\Lambda = 0.7, H_0 = 70$\} \\
SEDs for galaxy components & E (only elliptical galaxies) \\
\noalign{\smallskip}
\hline
\end{tabular}
}
\label{stuff}
\end{center}
$^{[1]}$ {\footnotesize Initial image is simulated for a large telescope so that the PSF effects are negligible.}
\end{table}

\begin{table}
\begin{center}
\caption{Galaxy Bins}
\begin{tabular}{lllr}
\hline\hline
\noalign{\smallskip}
Bin Name & Axis ratio &\multicolumn{2}{c}{Half-light Radius [arcsec]} \\
\noalign{\smallskip}
& & Mean & Variance \\
\noalign{\smallskip}
\hline
\noalign{\smallskip}
E1 & $<$0.3 & 0.2038 & 0.0014 \\
E2 & 0.3-0.4 & 0.1956 & 0.0010\\
E3 & 0.4-0.5 & 0.1915 & 0.0007\\
E4 & 0.5-0.6 & 0.1980 & 0.0022\\
E5 & 0.6-0.7 & 0.2063 & 0.0028\\
E6 & 0.7-0.8 & 0.1974 & 0.0010\\
E7 & 0.8-0.9 & 0.1922 & 0.0007\\
E8 & $>$0.9 & 0.2076 & 0.0042\\
\noalign{\smallskip}
\hline
\end{tabular}
\label{galaxyBins}
\end{center}
\end{table}

With the \textsc{stuff} catalog as input, FITS image of the galaxies are created using the tool \textsc{skymaker} \citep*{ber09}. The galaxies have a standard de--Vaucouleur profile. The de--Vaucouleur profile is fully defined by the centroid position (x,y), magnitude, effective radius, axis--ratio and position--angle of the galaxy. In principle, \textsc{skymaker} allows users to set telescope PSF and CCD detector characteristics, but we used it as an simple image generator. We set the telescope and detector characteristics such that the resulting images have negligible instrumental artifacts, pixelisation and noise. For computation speed, images are cut-out into 512$\times$512 pixels stamps centered on galaxies. We  convolved the stamps of images with PSFs (cf section 2.5) of both the designs. The convolved images have an infinite SNR. We add poissonian noise equivalent to sky background of 30 mag/arcsec$^2$ to the images to reduce their SNR to desired value. For real sky background \citep*{lei98} we need to increase the exposure time to obtain the desired SNR for both the designs. Noise addition is done for images with both pixel scales. The essential inputs for the \textsc{skymaker} tool are shown in Table \ref{skymaker}.

\begin{table}
\begin{center}
\caption{Essential input information for \textsc{skymaker} image simulator}
\resizebox{8cm}{!}{
\begin{tabular}{ll}
\hline\hline
\noalign{\smallskip}
Input parameter & Value \\
\noalign{\smallskip}
\hline
\noalign{\smallskip}
Exposure time (seconds) & 58$^{[1]}$ \\
Magnitude Zeropoint ($mag$) & 30.7 \\
Pixel size ($arcsec$) & 0.025 \\
M$_1$ \PM\, diameter ($m$) & 10 \\
Wavelength ($\mu m$) & 0.55 \\
Background surface brightness ($mag/arcsec^2$) & 50 $^{[2]}$ \\
\noalign{\smallskip}
\hline
\end{tabular}
}
\label{skymaker}
\end{center}
$^{[1]}$ {\footnotesize The Exposure time is equivalent to a 10m telescope. To obtain the same flux and SNR, a 1.2-m telescope should have an exposure time in the order of 4050 seconds \citep*{sch87}.} \\
$^{[2]}$ {\footnotesize The image simulated will contain no sky background noise. This will be added after PSF convolution.}
\end{table}

\subsection{Model Fitting Techniques}
We setup a two-step pipeline to fit galaxy models to the simulated ellipticals. First we extracted the galaxies and performed a crude measurement of axis--ratio and position--angle using the routine \textsc{sextractor} \citep*{ber96}. Second, we used the more elaborate  2D-parametric galaxy fitting routine \textsc{galfit} \citep*{pen10} with the \textsc{sextractor} derived morphological parameters as initial values. Advantages of using \textsc{galfit} are, it takes the PSF effects into account while fitting the parameters, it is not very sensitive to initial settings and \textsc{galfit} gives chi-squared values of the fit as an indication on the confidence. \textsc{galfit} optimizes a set of parameters over an input 2D-image. The user decides which parameters are kept constant and which ones are fitted. In addition to the 2D-image and the PSF, \textsc{galfit} requires the correct exposure time and magnitude zeropoint as input. \textsc{galfit} is especially sensitive to sky  background and allows users to fix the sky value or fit the best sky value. Since the galaxies are simulated using de--Vaucouleur profile, the same is fitted for the final image using \textsc{galfit}. The (x,y) position and the magnitude are fixed using the values obtained from \textsc{sextractor}. The values fitted by \textsc{galfit} are the effective radius, axis--ratio and position--angle of the galaxy. Now we compare the simulated and fitted values of the axis--ratio and position--angle. We ensure that all galaxies are fit with a reduced chi-squared close to one. The standard deviation allowed for the reduced chi-squared is 0.2. Galaxies fitted outside this regime are not included for further analysis. Since \textsc{galfit} uses the PSF as an input, we can use PSFs from other regions in the FoV to study the effect on precision due to PSF reconstruction errors.

\section{Results and Discussions}
In this section, we present the mean error and 1$\sigma$ standard deviation in the axis--ratio and position--angle (PA) measurement for both the designs. Later we discuss the results obtained. The mean error and 1$\sigma$ error bar are presented for 3 case studies. In case study~\#1 we assume it is possible to perfectly reconstruct the PSF for model fitting. In essence we use the same PSF for image simulation and model fitting. This is the best case scenario. In case study~\#2 we take into account the fact that it is not possible to perfectly reconstruct the PSF. So we use a PSF from a far region in the FoV for fitting process. This is the worst case scenario. In case study~\#3 we use a number of PSF between the best case and worst case scenario to mimic possible errors in realistic PSF reconstruction.

The difference between the simulated and the fitted values (of morphological properties) is the error in measurement. This error can be used to compare the image quality of the two designs. The error in axis-ratio and position-angle of the fitted values is measured for hundred galaxies in each axis-ratio bin (see Table \ref{galaxyBins}). If $q$ is the simulated axis-ratio of galaxy and $q'$ the fitted axis-ratio, the error in expressed by the difference $(q - q')/q$. Assuming that the error set (for 100 galaxies) has a normal distribution we calculate the mean and 1$\sigma$ standard deviation for the errors in each bin. The mean and especially the 1$\sigma$ standard deviation should be lower for the design with better image quality.

\subsection{Case \#1}

\begin{figure}
\begin{center}
\includegraphics[width = 3in, keepaspectratio=true]{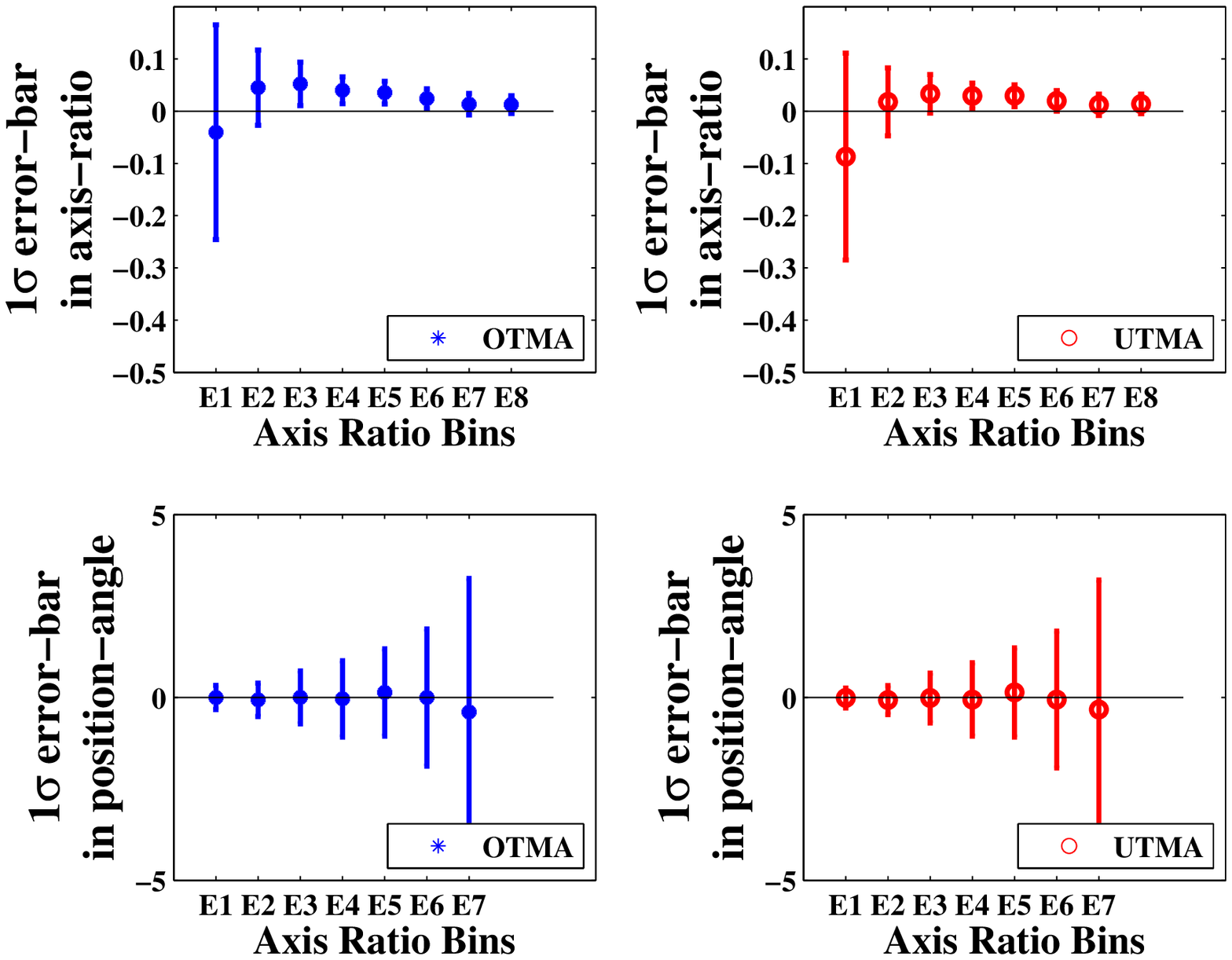}
\caption{{\bf Error-bar Plot:} The error-bar plot corresponding to case study \#1 for axis--ratio and position--angle measurement. The images have a pixel scale of 0.025\arcsec.}
\label{errorp025}
\includegraphics[width = 3in, keepaspectratio=true]{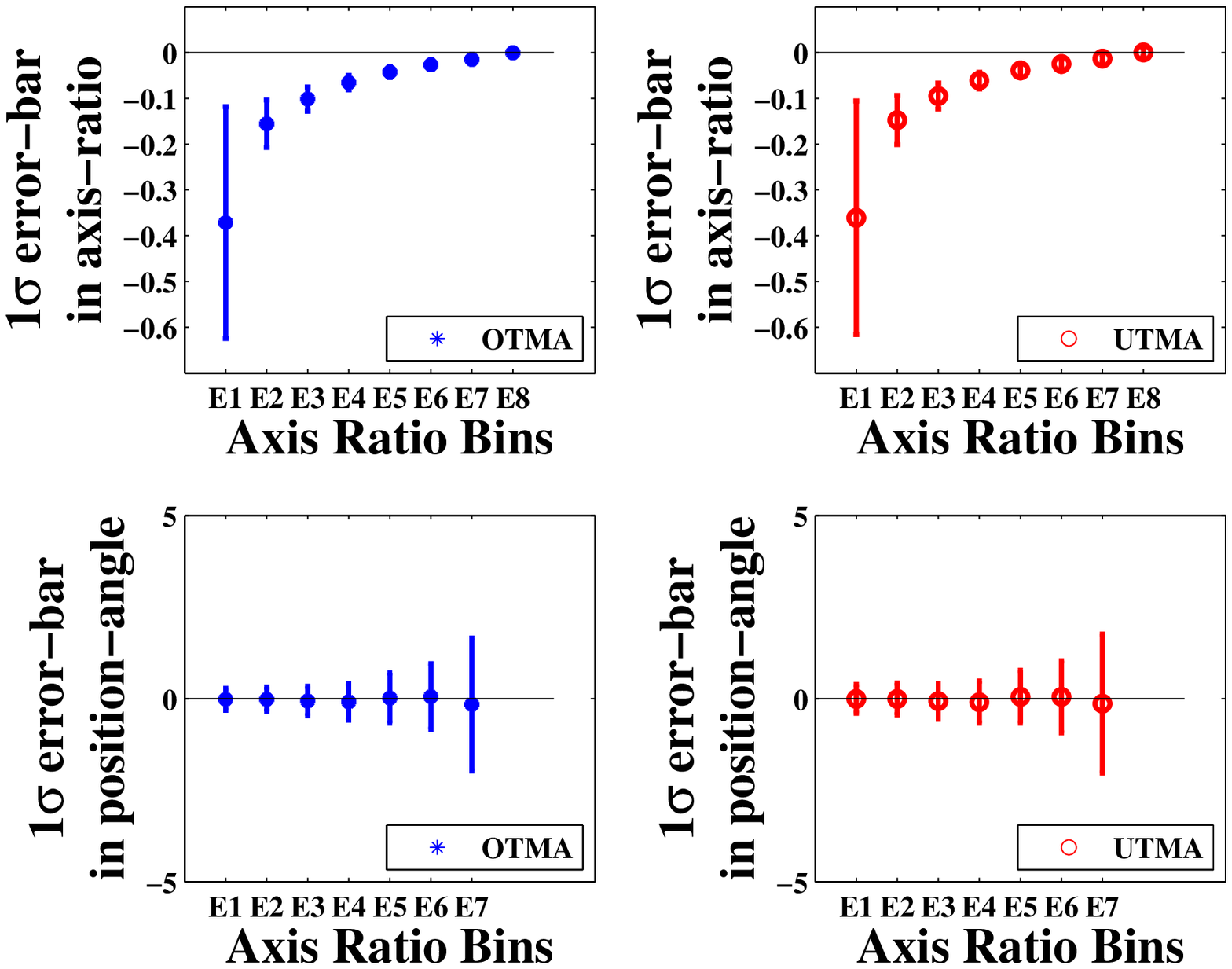}
\caption{{\bf Error-bar Plot:} The error-bar plot corresponding to case study \#1 for axis--ratio and position--angle measurement. The images have a pixel scale of 0.1\arcsec. The large pixel size introduces a systematic bias in the axis--ratio measurement for both the designs.}
\label{errorp1}
\end{center}
\end{figure}

In this case study we use the same PSF (computed at point labeled `1' in Figure \ref{tmaFoV}) for both simulation and fitting process. Tables \ref{caseStudy1q} and \ref{caseStudy1pa} present the mean error and 1$\sigma$ error-bar for axis--ratio and PA measurement for both pixel scales. Figures \ref{errorp025} and \ref{errorp1} show the error-bar plots for both the telescopes corresponding to images with pixel size $0.025\arcsec$ and $0.1\arcsec$ respectively. In both the Figures the x-axis runs from axis--ratio bin E1(q $<$ 0.3) through E8(q $>$0.9). 

Figure \ref{errorp025} shows the error-bar plot for axis--ratio and PA measurement for images with pixel scale $0.025\arcsec$. The results are almost same for both the OTMA and UTMA design. The mean error is close to zero for axis--ratio measurement but the 1$\sigma$ error-bar is high for slender galaxies and is negligibly small for circular galaxies. This is a consequence of the fact that the minor axis of the slender galaxies is not properly resolved by both the designs. In case of the PA measurement the mean error is close to zero for both the designs and the 1$\sigma$ error-bar is small for slender galaxies and increases as galaxies get circular. The PSF generally makes all objects circular and this affects the precision with which PA can be measured for intrinsically circular objects. The last bin E8(q $>$0.9) is not shown for PA plots because these objects are almost circular and there is no sense in measuring their PA.

Figure \ref{errorp1} shows the error-bar plot for axis--ratio and PA measurement for images with pixel scale $0.1\arcsec$. The results are again the same for both the OTMA and UTMA design. In the case of large pixels, for axis--ratio measurements the mean error is biased towards negative values. This implies that the axis--ratio is always over estimated. This over estimation can be attributed to the PSF effects, large pixel size and pixel shape. Contemporary missions are designed with pixel sizes of this order and this bias needs to be accounted for.

From these results we can conclude that if the PSF is accurately reconstructed, we can perform the science measurements with the same precision using both the designs.  The transfer of energy from the Airy disk to the outer rings or the diffraction spikes shown in Figure \ref{tmaPSF} for the OTMA design do not affect the science measurements of this design significantly. But in real time analyses there is always an uncertainty in the knowledge of the PSF.

\begin{table*}
\begin{minipage}{7in}
\begin{center}
\caption{The mean error in axis--ratio measurement and the 1$\sigma$ error-bar for the same are given for case study \#1}
\resizebox{16cm}{!}{
\begin{tabular}{lllllllll}
Axis--ratio bin & \multicolumn{4}{c}{Pixel Scale 0.025$\arcsec$} & \multicolumn{4}{c}{Pixel Scale 0.1$\arcsec$} \\
\noalign{\smallskip}
\hline
\noalign{\smallskip}
&\multicolumn{2}{c}{OTMA} & \multicolumn{2}{c}{UTMA} & \multicolumn{2}{c}{OTMA} & \multicolumn{2}{c}{UTMA} \\
\noalign{\smallskip}
\hline
\noalign{\smallskip}
& Mean error &  1$\sigma$ error-bar & Mean error &  1$\sigma$ error-bar & Mean error &  1$\sigma$ error-bar & Mean error &  1$\sigma$ error-bar\\
\noalign{\smallskip}
\hline
\noalign{\smallskip}
E1 & -0.0404 & 0.2056 & -0.0869 & 0.1981 & -0.3712 & 0.253 & -0.3609 & 0.2549 \\
E2 & 0.045 & 0.0721 & 0.0177 & 0.0648 & -0.1556 & 0.0519 & -0.1474 & 0.0536 \\
E3 & 0.0523 & 0.0416 & 0.0333 & 0.0367 & -0.1016 & 0.0266 & -0.0949 & 0.0282 \\
E4 & 0.0401 & 0.0257 & 0.0292 & 0.0245 & -0.0656 & 0.0155 & -0.0612 & 0.0181 \\
E5 & 0.0355 & 0.0216 & 0.0296 & 0.0208 & -0.0425 & 0.0121 & -0.0389 & 0.0128 \\
E6 & 0.024 & 0.0189 & 0.0198 & 0.0197 & -0.027 & 0.0107 & -0.0249 & 0.0109 \\
E7 & 0.0133 & 0.0206 & 0.0121 & 0.0203 & -0.0151 & 0.0091 & -0.0135 & 0.009 \\
E8 & 0.0127 & 0.017 & 0.0138 & 0.0187 & -0.0004 & 0.0097 & 0.0004 & 0.0102 \\
\noalign{\smallskip}
\hline
\end{tabular}
}
\label{caseStudy1q}
\end{center}
\end{minipage}
\end{table*}

\begin{table*}
\begin{minipage}{7in}
\begin{center}
\caption{The mean error in position--angle measurement and the 1$\sigma$ error-bar for the same are given for case study \#1}
\resizebox{16cm}{!}{
\begin{tabular}{lllllllll}
Axis--ratio bin & \multicolumn{4}{c}{Pixel Scale 0.025$\arcsec$} & \multicolumn{4}{c}{Pixel Scale 0.1$\arcsec$} \\
\noalign{\smallskip}
\hline
\noalign{\smallskip}
&\multicolumn{2}{c}{OTMA} & \multicolumn{2}{c}{UTMA} & \multicolumn{2}{c}{OTMA} & \multicolumn{2}{c}{UTMA} \\
\noalign{\smallskip}
\hline
\noalign{\smallskip}
& Mean error &  1$\sigma$ error-bar & Mean error &  1$\sigma$ error-bar & Mean error &  1$\sigma$ error-bar & Mean error &  1$\sigma$ error-bar\\
\noalign{\smallskip}
\hline
\noalign{\smallskip}
E1 & -0.0006 & 0.3289 & -0.0165 & 0.2709 & -0.0181 & 0.2982 & -0.0038 & 0.3939 \\
E2 & -0.065 & 0.4563 & -0.0697 & 0.3959 & -0.015 & 0.3288 & -0.0056 & 0.4352 \\
E3 & 0.0046 & 0.7227 & -0.0167 & 0.6791 & -0.0597 & 0.4013 & -0.0691 & 0.4913 \\
E4 & -0.0389 & 1.0421 & -0.0569 & 1.0044 & -0.0845 & 0.4966 & -0.0938 & 0.5733 \\
E5 & 0.1375 & 1.1953 & 0.1358 & 1.2191 & 0.0195 & 0.6845 & 0.0534 & 0.7226 \\
E6 & -0.0004 & 1.8753 & -0.058 & 1.8711 & 0.0575 & 0.896 & 0.0507 & 0.9819 \\
E7 & -0.3974 & 3.6504 & -0.33 & 3.5373 & -0.1562 & 1.8145 & -0.1337 & 1.8951 \\
E8 & -0.3685 & 21.4918 & -1.0727 & 27.5538 & -1.4601 & 16.4343 & 0.9883 & 19.3395 \\
\noalign{\smallskip}
\hline
\end{tabular}
}
\label{caseStudy1pa}
\end{center}
\end{minipage}
\end{table*}

\subsection{Case \#2}
In this case study we use a PSF from a different region in the FoV (computed at point labeled `11' in the Figure \ref{tmaFoV}) for model fitting purposes. This is equivalent to having poor knowledge of the PSF.

\begin{figure}
\begin{center}
\includegraphics[width = 3in, keepaspectratio=true]{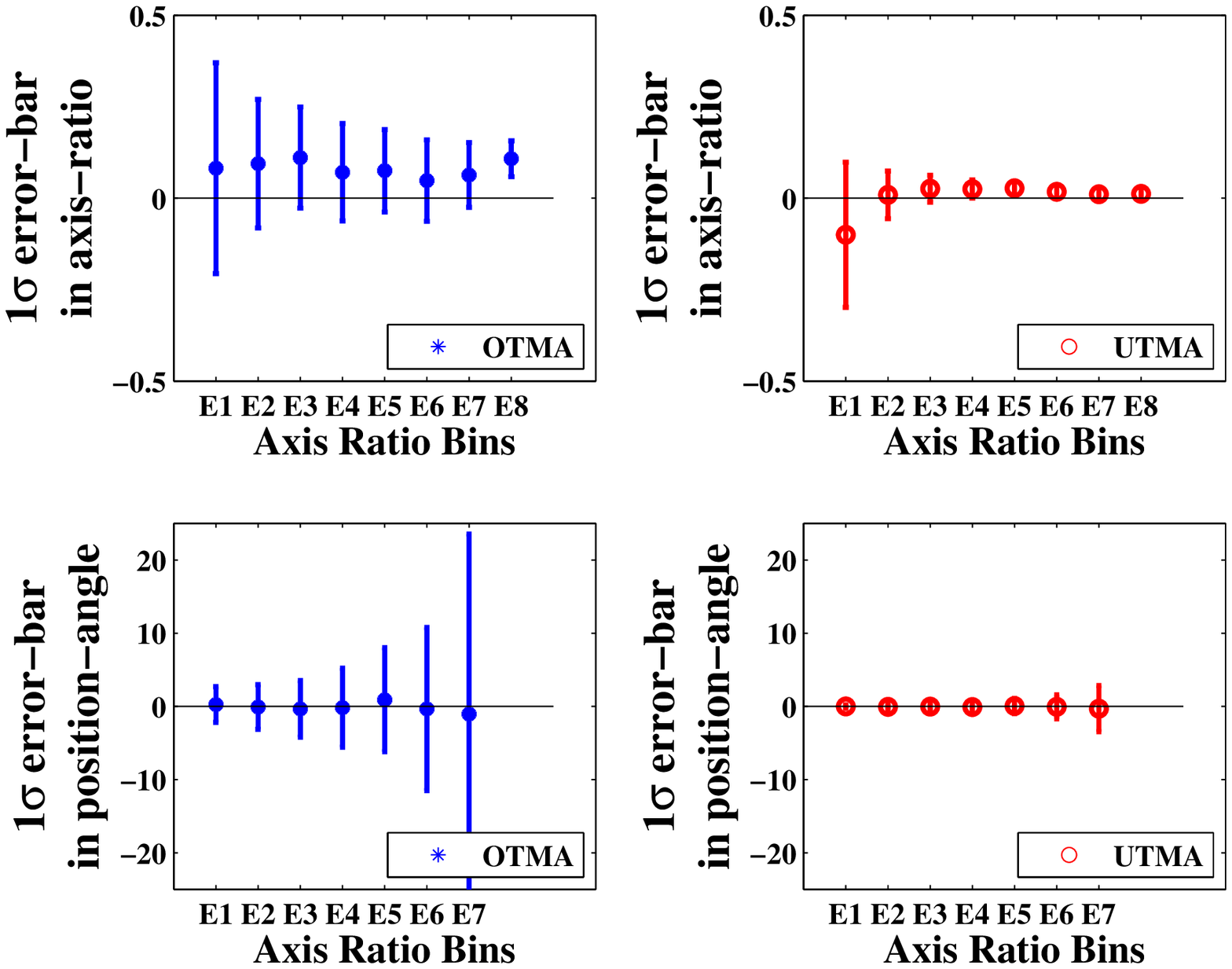}
\caption{{\bf Error-bar Plot:} The error-bar plot corresponding to case study \#2 for axis ratio and position angle measurement. The images have a pixel scale of 0.025\arcsec.}
\label{errorp025w}
\includegraphics[width = 3in, keepaspectratio=true]{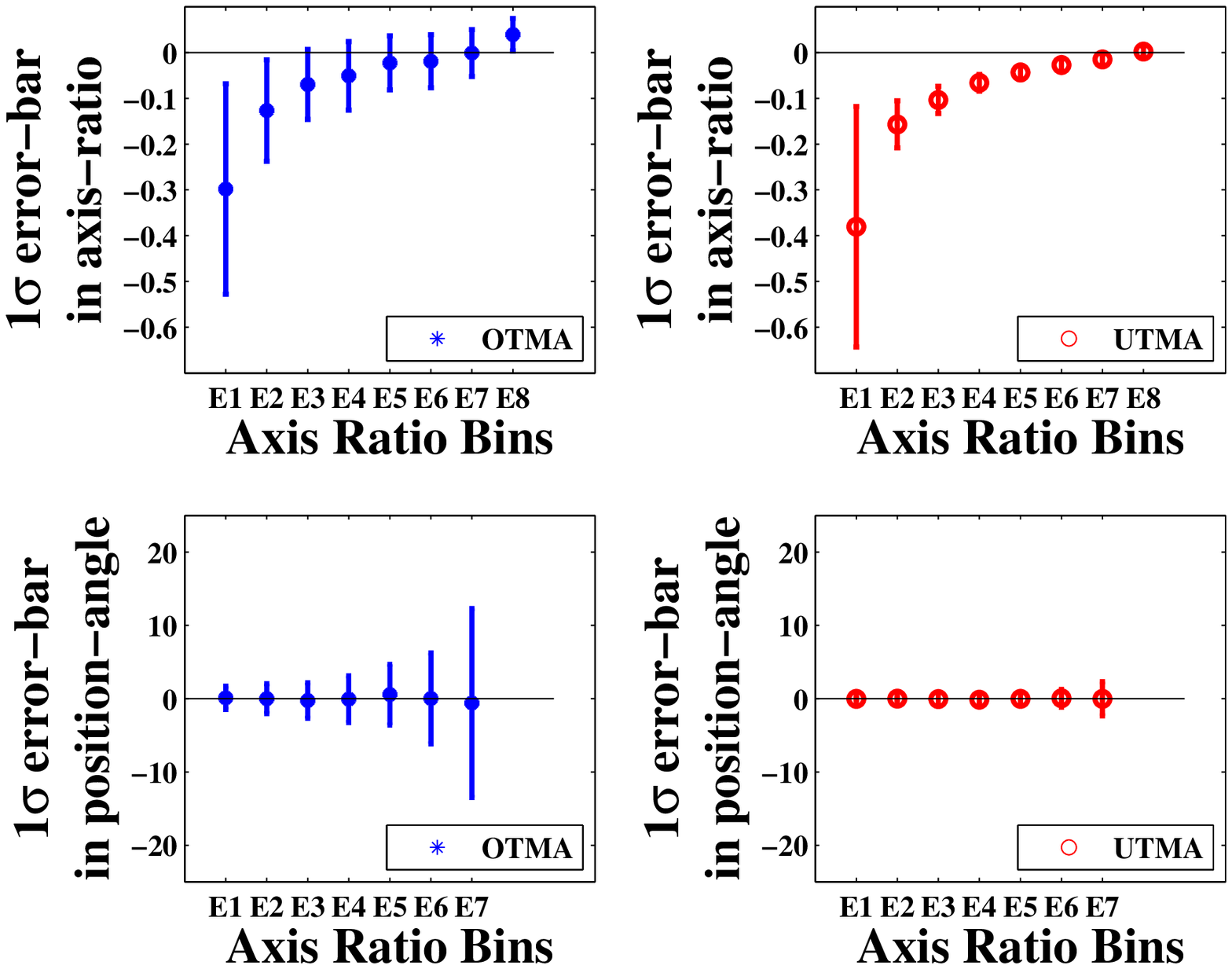}
\caption{{\bf Error-bar Plot:} The error-bar plot corresponding to case study \#2 for axis ratio and position angle measurement. The images have a pixel scale of 0.1\arcsec. The large pixel size introduces a systematic bias in the axis--ratio measurement for both the designs.}
\label{errorp1w}
\end{center}
\end{figure}

The error-bars in axis--ratio and PA measurement are presented in Figures \ref{errorp025w} and \ref{errorp1w} for pixel scales $0.025\arcsec$ and $0.1\arcsec$ respectively. Tables \ref{caseStudy2q} and \ref{caseStudy2pa} present the mean error and 1$\sigma$ error-bar for axis--ratio and PA measurement for both pixel scales. For images with pixel scale $0.025\arcsec$ the 1$\sigma$ error-bar is smaller for the UTMA design than the OTMA design for both the axis--ratio and PA measurement. The error bars are smaller by a factor of 2-4 for the UTMA depending upon the axis--ratio bin. In other words, even in the presence of PSF reconstruction errors the 1$\sigma$ error-bars of the UTMA design do not change, but that of the OTMA increase in size for both the axis--ratio and PA measurement. The same results are obtained for pixel size $0.1\arcsec$ too.

Since the UTMA PSF does not vary over the FoV, all the unsaturated point sources in the FoV can be used to extract the PSF. Hence, it is possible to obtain high SNR PSFs for the UTMA design.

\begin{table*}
\begin{minipage}{7in}
\begin{center}
\caption{The mean error in axis--ratio measurement and the 1$\sigma$ error-bar for the same are given for case study \#2}
\resizebox{16cm}{!}{
\begin{tabular}{lllllllll}
Axis--ratio bin & \multicolumn{4}{c}{Pixel Scale 0.025$\arcsec$} & \multicolumn{4}{c}{Pixel Scale 0.1$\arcsec$} \\
\noalign{\smallskip}
\hline
\noalign{\smallskip}
&\multicolumn{2}{c}{OTMA} & \multicolumn{2}{c}{UTMA} & \multicolumn{2}{c}{OTMA} & \multicolumn{2}{c}{UTMA} \\
\noalign{\smallskip}
\hline
\noalign{\smallskip}
& Mean error &  1$\sigma$ error-bar & Mean error &  1$\sigma$ error-bar & Mean error &  1$\sigma$ error-bar & Mean error &  1$\sigma$ error-bar\\
\noalign{\smallskip}
\hline
\noalign{\smallskip}
E1 & 0.082 & 0.288 & -0.1001 & 0.1985 & -0.2982 & 0.2297 & -0.3805 & 0.2628 \\
E2 & 0.0944 & 0.176 & 0.009 & 0.065 & -0.1267 & 0.1108 & -0.1569 & 0.0513 \\
E3 & 0.1109 & 0.1383 & 0.026 & 0.0369 & -0.0696 & 0.0765 & -0.1037 & 0.0296 \\
E4 & 0.0709 & 0.1332 & 0.0251 & 0.0245 & -0.051 & 0.0749 & -0.0664 & 0.0181 \\
E5 & 0.075 & 0.1127 & 0.0269 & 0.0195 & -0.0226 & 0.0591 & -0.0438 & 0.0154 \\
E6 & 0.0481 & 0.1114 & 0.0177 & 0.018 & -0.0191 & 0.0578 & -0.0275 & 0.0143 \\
E7 & 0.0636 & 0.0885 & 0.0109 & 0.0183 & -0.0011 & 0.0512 & -0.0149 & 0.0118 \\
E8 & 0.1078 & 0.049 & 0.012 & 0.0175 & 0.039 & 0.0354 & 0.0022 & 0.0115 \\
\noalign{\smallskip}
\hline
\end{tabular}
}
\label{caseStudy2q}
\end{center}
\end{minipage}
\end{table*}

\begin{table*}
\begin{minipage}{7in}
\begin{center}
\caption{The mean error in position--angle measurement and the 1$\sigma$ error-bar for the same are given for case study \#2}
\resizebox{16cm}{!}{
\begin{tabular}{lllllllll}
Axis--ratio bin & \multicolumn{4}{c}{Pixel Scale 0.025$\arcsec$} & \multicolumn{4}{c}{Pixel Scale 0.1$\arcsec$} \\
\noalign{\smallskip}
\hline
\noalign{\smallskip}
&\multicolumn{2}{c}{OTMA} & \multicolumn{2}{c}{UTMA} & \multicolumn{2}{c}{OTMA} & \multicolumn{2}{c}{UTMA} \\
\noalign{\smallskip}
\hline
\noalign{\smallskip}
& Mean error &  1$\sigma$ error-bar & Mean error &  1$\sigma$ error-bar & Mean error &  1$\sigma$ error-bar & Mean error &  1$\sigma$ error-bar\\
\noalign{\smallskip}
\hline
\noalign{\smallskip}
E1 & 0.2494 & 2.4463 & -0.0195 & 0.2412 & 0.1196 & 1.578 & -0.0173 & 0.4011 \\
E2 & -0.1053 & 3.0449 & -0.072 & 0.3546 & -0.0004 & 2.0329 & -0.0018 & 0.4576 \\
E3 & -0.3525 & 3.8699 & -0.0466 & 0.5481 & -0.2501 & 2.4371 & -0.0559 & 0.5421 \\
E4 & -0.1876 & 5.386 & -0.0899 & 0.8272 & -0.0634 & 3.1799 & -0.1272 & 0.657 \\
E5 & 0.8968 & 7.107 & 0.0479 & 1.0383 & 0.552 & 4.1362 & -0.0234 & 0.935 \\
E6 & -0.3629 & 11.1505 & -0.0641 & 1.64 & 0.042 & 6.1979 & 0.0555 & 1.2076 \\
E7 & -1.0651 & 24.6023 & -0.3125 & 3.1533 & -0.5958 & 12.8972 & -0.0202 & 2.3305 \\
E8 & -3.7607 & 46.6307 & -0.8581 & 28.2858 & -3.7111 & 42.5253 & -5.7275 & 23.0076 \\
\noalign{\smallskip}
\hline
\noalign{\smallskip}
\end{tabular}
}
\label{caseStudy2pa}
\end{center}
\end{minipage}
\end{table*}

\subsection{Case \#3}
\begin{figure}
\begin{center}
\includegraphics[width = 3in, keepaspectratio=true]{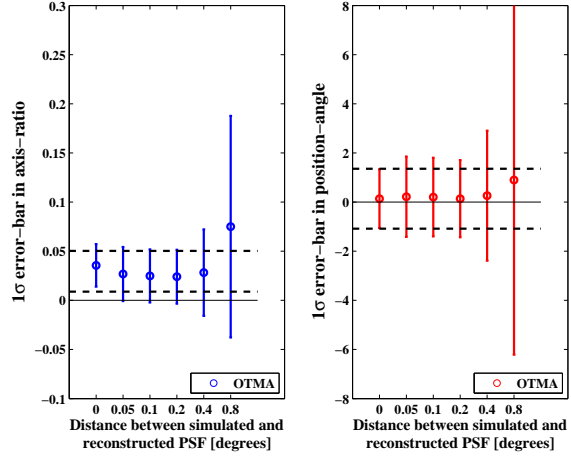}
\caption{{\bf Error-bar Plot:} The error-bar plot corresponding to case study \#3 for axis ratio and position angle measurement. The images have a pixel scale of 0.025\arcsec. The dotted bounding line is the worst case error-bar for UTMA design.}
\label{errorCase3}
\end{center}
\end{figure}

 In this case study we focus only on the OTMA design, because we showed in case study \#2 that the UTMA PSF is uniform over a wide region of the FoV. We sample OTMA PSFs at intermediate points between the simulated PSF ( labeled `1' in the Figure \ref{tmaFoV}) and worst case ( labeled `11' in the Figure \ref{tmaFoV}). The PSFs are sampled at 4 positions \{0.05,0.1,0.2,0.4\} degrees from the PSF used for simulation. For these PSFs we find the precision with which the OTMA design performs measurements for a particular axis--ratio bin (E5).  Figure \ref{errorCase3} shows the  mean error and 1$\sigma$ error-bar for axis--ratio and PA measurement for the intermediate PSFs. We find that the precision of the OTMA design is equivalent to that of the UTMA design when the PSF of the OTMA design can be reconstructed within 0.2 degrees or 12 arc-minutes from the simulated PSF. The error-bar for UTMA design case study \#2 is also shown for comparison.

\begin{table*}
\begin{center}
\caption{The mean error and the 1$\sigma$ error-bar for axis--ratio and position--angle measurement for case study \#3}
\begin{tabular}{lllll}
PSF distance$^*$  & \multicolumn{2}{c}{Axis--ratio} & \multicolumn{2}{c}{Position--Angle} \\
\noalign{\smallskip}
\hline
\noalign{\smallskip}
[in degrees] & Mean error &  1$\sigma$ error-bar & Mean error &  1$\sigma$ error-bar\\
\noalign{\smallskip}
\hline
\noalign{\smallskip}
0.0 & 0.0355 & 0.0216 & 0.1374 & 1.1953 \\
0.05 & 0.0267 & 0.0275 & 0.2167 & 1.6345 \\
0.1 & 0.0248 & 0.0269 & 0.2018 & 1.6002 \\
0.2 & 0.0240 & 0.0273 & 0.1377 & 1.5710 \\
0.4 & 0.0282 & 0.0439 & 0.2596 & 2.6446 \\
0.8 & 0.07496 & 0.1127 & 0.8968 & 7.1069 \\
\noalign{\smallskip}
\hline
\noalign{\smallskip}
\end{tabular}
\label{caseStudy3}
\end{center}
$^*$ {\footnotesize Distance between the simulated PSF and reconstruted PSF}
\end{table*}

\subsection{Discussion of the Results}

\subsubsection{Encircled Energy Plots}
\begin{figure}
\begin{center}
\includegraphics[width = 3in, keepaspectratio=true]{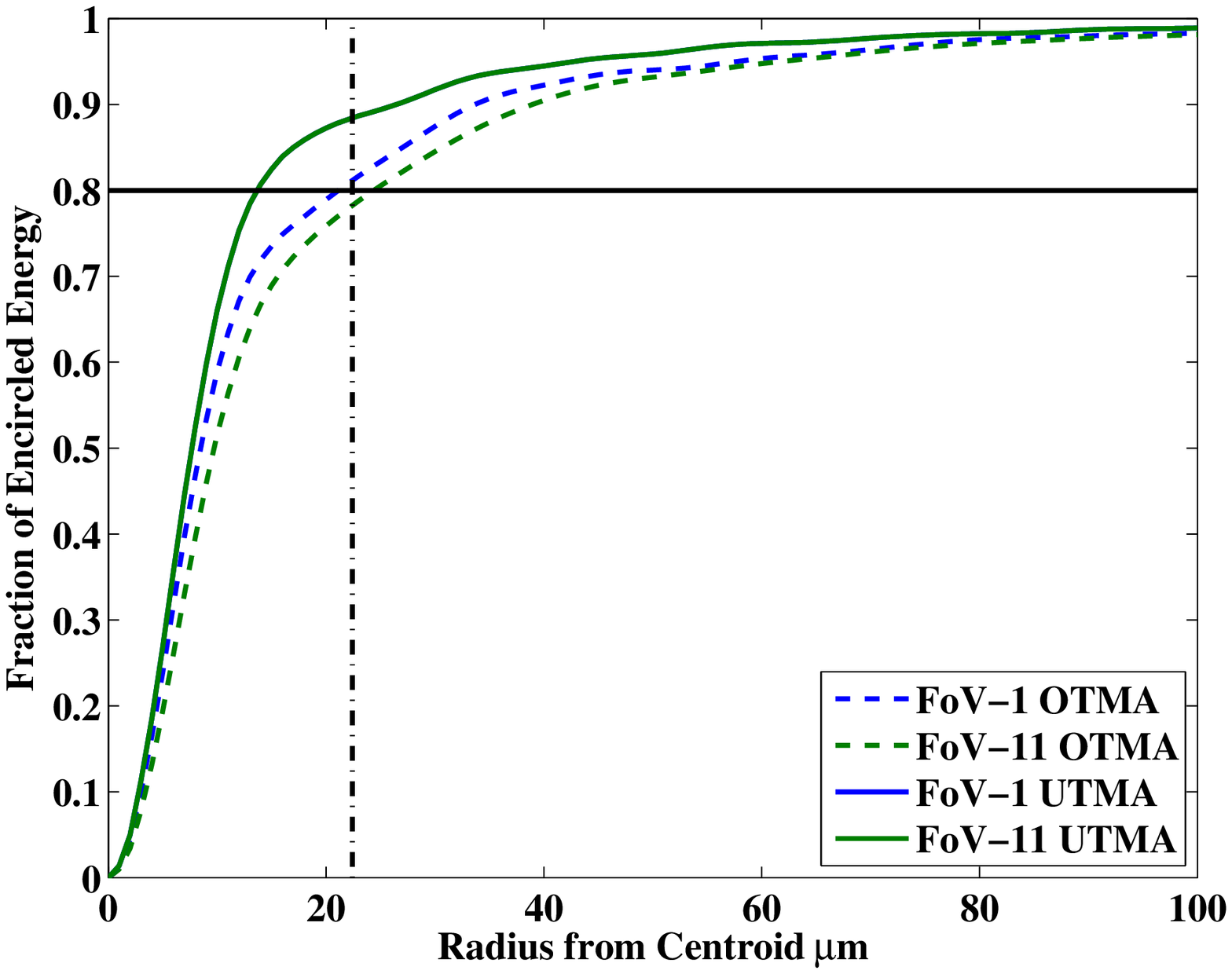}
\caption{{\bf EE plot:} The EE plot is shown for the OTMA in the left panel and the UTMA in the right panel. The EE is calculated for wavelengths between 550 nm and 900 nm at two different FoV locations. The labels for the FoV correspond to the labels in Figure \ref{tmaFoV}. \citet*{lam10} have shown that the radius for EE $=80\%$ level is greater for the OTMA design. We also show EE plot does not change for the UTMA design (solid line) over the FoV and it changes considerably for the OTMA design (dashed line).}
\label{EE}
\end{center}
\end{figure}

From the above case studies we conclude that, if we can reconstruct the PSF within 10 arc-minutes from the point of interest, the OTMA design perform science measurements with the same precision as UTMA design. The performance of the OTMA design drops if the PSF is reconstructed at a distance greater than 10 arc-minutes from the source. The reason can be understood if we look at the encircled energy (EE) plots at different regions in the FoV for both the designs.  Figure \ref{EE} shows the EE plot at 2 different regions in the FoV for both the designs. These 2 regions correspond to the region of interest in the above case studies (labeled FoV-1 and FoV-11 in Figure \ref{tmaFoV}). The EE varies over the FoV for the OTMA design but it is uniform for the UTMA design.  Since EE is a derivative of the PSF we can conclude that the PSF of the UTMA design is homogeneous over the entire FoV. The same is not true for the OTMA design.

The OTMA design used here is only a representative of a family of OTMA designs. Therefore, one can argue that it is possible to design other OTMA telescopes that have uniformly same EE over the entire FoV. We show here that the presence of spider obstruction will not allow OTMA designs to have homogenous PSF over a wide FoV.  We modeled the OTMA1 telescope in \textsc{zemax} without the spider support structure and computed the EE at different FoV locations. The EE plot of the design without spider support is reasonably uniform over the entire FoV. Therefore, the spider support structure is the significant contributor for the inhomogeneity to the PSF. Hence, in theory the UTMA PSFs are homogenous over a wider FoV compared to OTMA PSFs.

\subsubsection{PSF Reconstruction}
For an OTMA design, the correlation between neighboring PSFs can be used to construct a metric which tell us how the PSF varies over the FoV. If sufficient number of unsaturated point sources are available in the FoV, the PSF can be computed at any point with desired SNR for an OTMA design. In case of UTMA design the PSF is uniform over the FoV and hence the PSF can be extracted without extensive knowledge of the metric. In practice the PSF for both the designs can be extracted with desired SNR for a given number of unsaturated point sources. If the PSF is known with desirable SNR, then both the designs will measure the morphological parameters with same precision as shown in case study \#1.

\subsubsection{Systematic Errors}
A short list of the important systematic errors that affect both the designs is as follows. Studies of the photometric evidence by \citet*{kin77} suggests that axis--ratio and position--angle change with isophotal radius in elliptical galaxies due to their triaxial nature. Also studies by \citet*{voi12} show that the impact of color gradients, which are intrinsically present on the image can affect the morphology measurements. Studies by \citet*{rho10} show that CTI effects are a dominant factor which affect the precision of aging CCDs. 

\subsubsection{Tolerancing}
Tolerancing or sensitivity analysis computes the change in a given property for a change in the constructional parameters. We do not perform a full scale tolerancing analysis to measure the change in the science results for tolerances in all the constructional parameters. Instead of such an exhaustive analysis, we did some defocussing to see how it affects the RMS wavefront error in the system. For reasonable defocussing of $\pm$ 0.01 mm, both the designs show an RMS wavefront error less than 0.07 waves, which is the diffraction limit of the system. The sensitivity of  all the constructional parameters needs to be tested include the manufacturing and alignment errors which will affect the PSF computed in this study. 

\section{Conclusion}
The significant results of our analysis are as follows. In case study \#1 and \#3 we show that the intrinsic properties of the OTMA PSF, like the bright concentric rings and diffraction spikes did not affect the morphological parameter measurements significantly, if the PSF is reconstructed within 10 arc-minutes of the source. If the PSF is reconstructed beyond 10 arc-minute the performance of the OTMA design degrades. But case study \#2 shows that the precision of the UTMA design is not affected even if the PSF is reconstructed 50 arc-minutes from the point of interest. This is because the PSF of  UTMA design is homogeneous over the desired FoV compared to the OTMA design.

\begin{acknowledgements}
We thank the referee for all the insightful remarks. These remarks helped us to strengthen and improve our arguments. We also thank Genevieve Soucail for providing valuable comments on the final draft of the manuscript. Author Balasubramanian S wishes to acknowledge ASTRIUM for a thesis fellowship that made this work possible.
\end{acknowledgements}

%-------------------------------------------------------------------

\appendix

\section[]{Constraints chosen to obtain a family of fundamental design parameters for OTMA and UTMA telescopes}
Any TMA telescope is defined by it fundamental design parameters explained in section 2.2. Using the fundamental design parameters one can obtain the constructional parameters using the relations derived in \cite{rob78} (See Table \ref{tmaConstEqn}). Mission requirements constrained only 3 of these fundamental parameters \{\Yone,\fthree,$\bar{u}_0$\} . Our task is to chose the best values for the remaining parameters \{\fpri, \ftwo, B, \Dthree\}. There are several possible values possible for these parameters. Small values of \fpri, and large values of \done\, keep the overall design shorter and the secondary aperture smaller. The upper limit on \done\, is $d_1 < F_p$ where $F_p = 2Y_1$\fpri. Best values for \done\, lie between 0.8\Fp\, and 0.9\Fp.  Cassegranian designs are compact so $F_2>0$ is desirable. In addition if $B>0$, the TM is aft of the \PM\, vertex and hence free of spurious reflections. Finally \Dthree\, can be constrained using the Petzval condition for flat focal plane \citep{rob78}.

\begin{table}
\begin{center}
\caption{Constructional parameters for TMA design expressed in terms of the fundamental parameters}
\begin{tabular}{lll}
\hline\hline
Surface\# & Curvature & Distance to next surface \\
\hline
1 Primary & $c_1 = \frac{-1}{2F_p}$ & $d_1 = -\frac{F_2 - B}{1+A_2}$ \\
2 Secondary & $c_2 = -\frac{1 - A_2}{2F_2(2d_1c_1 - 1)}$ & $d_2 = -d_1 + B + D_3$ \\
3 Tertiary & $c_3 = \frac{-1}{2D_3}[1-\frac{S_0S_1F_2}{F_3}]$ & $d_3 = \frac{\|D_3\|S_1F_3}{F_2}$ \\
4 Image & $c_4 = -2(c_1-c_2+c_3)$ & na \\
\hline
\end{tabular}
\label{tmaConstEqn}
\end{center}
Focal length of the primary mirror $F_p = 2Y_1F_{pri}$ \\
Focal length amplification of the two mirror system $A_2 = \frac{F_2}{F_p}$ \\
Logical variables defined $S_0 = \frac{\|D_3\|}{D_3}$ and $S_1 = -\frac{\|F_2\|}{F_2}$ \\
For the equations to obtain the conic constants see \cite{rob78}.
\end{table}

Choose a range of values for \fpri. This constraints the allowable values for \done\, between 0.8\Fp $<$\done$<$ 0.9\Fp. Using the relations in Table \ref{tmaConstEqn} the minimum value of \ftwo\, that satisifies $F_2>0$ and $B>0$ can be found. It is given by Equation. And the maximum value for \ftwo\, is arbitrarily set at 2\fthree. 
\begin{align}
F_{2min} = \frac{-d_1}{1+d_1/F_p}
\end{align}
We have three parameters which can take a range of values \{\fpri,\done,\ftwo\}. The value of B can be obtained from the relations in Table \ref{tmaConstEqn}. The value of \Dthree\, can be found using the Petzval condition for flat focal plane. There are two possible solution for \Dthree, positive or negative. The positive value is chosen, so that the the design can be folded. The range of \fpri\, values chosen for the \otma\, design lie between 0.5 and 1.5. And for the \utma\, design it is 4.5-5.5. Using the equations above a large set of \done, \ftwo\, B and \Dthree\, are obtained. The fundamental parameters are then used to obtain the constructional parameters. A few designs are randomly chosen from the large set and modeled in the optical design program \textsc{zemax} to study their characteristics.

\end{document}